\documentclass[11pt,twoside]{article}
\usepackage{graphicx}
\usepackage{hyperref}

\begin{document}

\begin{center}
{\Large {\bf Fast single-dish scans of the Sun using ALMA}}

\bigskip
\textit{Neil Phillips$^{a}$, Richard Hills$^b$, Tim Bastian$^c$, \\
Hugh Hudson$^d$, Ralph Marson$^c$, and Sven Wedemeyer$^e$}
\footnote{
$^a$Joint ALMA Observatory;
$^b$University of Cambridge, UK;
$^c$National Radio Astronomy Observatory, Socorro NM, USA;
$^d$University of California, Berkeley CA, USA;
$^e$Institute of Theoretical Astrophysics, University of Oslo, Norway.}
\end{center}

\bigskip
%
%\author{Sample~Author1,$^1$ Sample~Author2,$^2$ and Sample~Author3$^3$
%\affil{$^1$Institution Name, Institution City, State/Province, Country; \email{AuthorEmail@email.edu}}
%\affil{$^2$Institution Name, Institution City, State/Province, Country; \email{AuthorEmail@email.edu}}
%\affil{$^3$Institution Name, Institution City, State/Province, Country; \email{AuthorEmail@email.edu}}}

% This section is for ADS Processing.  There must be one line per author.
%\paperauthor{Sven~Wedemeyer}{sven.wedemeyer@astro.uio.no}{http://orcid.org/0000-0002-5006-7540}{University of Oslo}{Institute of Theoretical Astrophysics}{Oslo}{}{0315}{Norway}
%\paperauthor{Asbjorn~Parmer}{}{}{University of Oslo}{Institute of Theoretical Astrophysics}{Oslo}{}{0315}{Norway}

We have implemented control and data-taking software that makes it
possible to scan the beams of individual ALMA antennas to perform
quite complex patterns while recording the signals at high rates. 
We conducted test observations of the Sun in September and December, 2014.
The data returned have excellent quality; in particular they allow us to
characterize the noise and signal fluctuations present in this kind of
observation. 
The fast-scan experiments included both Lissajous
patterns covering rectangular areas, and ``double-circle'' patterns of
the whole disk of the Sun and smaller repeated maps of specific disk-shaped targets. 
With the latter we find that we can achieve
roughly Nyquist sampling of the Band~6 (230~GHz) beam in 60~s over
a region 300$''$ in diameter. 
These maps show a peak-to-peak
brightness-temperature range of up to 1000~K, while the time-series
variability at any given point appears to be of order 0.5\% RMS over times of a few minutes. 
We thus expect to be able to separate the noise contributions due to
transparency fluctuations from variations in the Sun itself. 
Such timeseries have many advantages, in spite of the non-interferometric observations.
In particular such data should make it possible to observe microflares in active regions
and nanoflares in any part of the solar disk and low corona.

\begin{figure}[htbp]
\centering
   \includegraphics[width=0.6\textwidth]{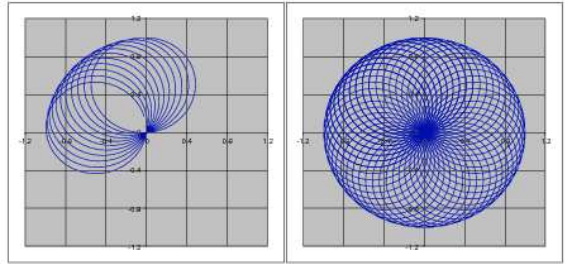}
      \caption{A double-circle scan pattern (left, just beginning; right, completed) composed of 2:1 amplitude ratio and fast scanning on the small circle.
      }  
\label{fig:doublecircle}
\end{figure}

The ALMA dishes permit solar data to be obtained on the fly, as an antenna executes a smooth but quite rapid motion.
Precise tracking corrections allow the antennas to be driven at frequencies of order 1~Hz, with excellent SNR at
sampling times of msec.
Area scanning with a pencil beam, for example with a Lissajous pattern, has a long heritage in radio astronomy
(e.g., Kovacs et al. 2008).
A Lissajous drive generally concentrates the coverage on the boundaries of a rectangular area; we have also used
``double circle'' drive patterns, the sum of large and small drive circular functions (Figure~\ref{fig:doublecircle} shows a double-circle pattern with the functions in a 2:1 amplitude ratio).
These have the merit of frequent return to the pattern center, thus providing relatively high-frequency sampling of the same part of the Sun.
Since the solar image varies only slowly, outside of flares, this capability acts as a guide to the system's radiometric stability of the entire pattern at a frequency of order 1~Hz.
This in some measure replaces chopping for this purpose.

%\articlefigure[width=.5\textwidth]{hughhudson_fig1.eps}{fig:doublecircle}{A double-circle scan pattern (left, just beginning; right, completed) composed of 2:1 amplitude ratio and fast scanning on the small circle.}

The fast scanning of a small area, centered on an active region is an excellent way to achieve mapping observations with high time resolution over an area larger than the beam width.
Given the stability of the observations (see below) this will enable ALMA to probe radiometric parameter space far better than any previous system, even with the use of just a single dish: such data will yield definitive observations of microflares and nanoflares 
(e.g., Hudson 1991).

\begin{figure}[htbp]
\centering
   \includegraphics[width=0.6\textwidth]{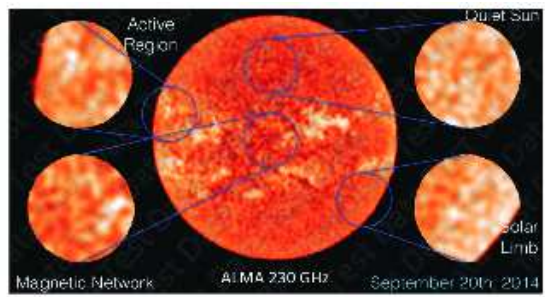}
      \caption{Images made from double-circle scans of the full Sun and of four sub-regions 300$''$ in diameter.
      The wisps above the solar limb (upper left small image) are real solar features.
      }  
\label{fig:testimages}
\end{figure}

Figure~\ref{fig:testimages} shows maps made from large and small double-circle scans with a single dish (PM01) at the ALMA low site during tests in September 2014).

%\articlefigure[width=.8\textwidth]{hughhudson_fig2.eps}{fig:testimages}{Test images from the September 2014 campaign, using double-circle scanning over the whole Sun and four sub-regions.
%The wisps above the solar limb (upper left small image) are real solar features.}

The structure on arcmin scales seen in all five of the images in Figure~\ref{fig:testimages} reflects the presence of the chromospheric network, which the ALMA test data show to be highly reproducible on 60-s image cadences.
This hints at the unprecedented accuracy and precision we believe ALMA will bring to basic subjects such as limb brightening.

\bigskip
{\bf Acknowledgements.}
We thank the Alma Solar Development Team and the ALMA Science Team for generous support during the test observations. 

%\bibliographystyle{asp2014}

%\bibliography{fastscan}  

% For BibTex

%[1] Wedemeyer-Bšhm, Scullion, Steiner et al. 2012, Nature, 486, 505
%[2] Freytag, Steffen, Ludwig, Wedemeyer-Bšhm et al. 2012, JCoPh, 231, 919 [3] Steffen, Ludwig, Wedemeyer. http://www.aip.de/~mst/linfor3D_main.html [4] Wedemeyer-Bšhm, Ludwig, Steffen, et al. 2007, A&A, 471, 977
%
%% For non-BibTex:
%\begin{thebibliography}{}

%\bibitem[{{Hudson}(1991)}]{1991SoPh..133..357H}
\bigskip
\begin{center}
{\bf References}
\end{center}

\bigskip
{Hudson}, H.~S. 1991, Solar Phys., 133, 357.

%\bibitem[{{Kov{\'a}cs}(2008)}]{2008SPIE.7020E...5K}
{Kov{\'a}cs}, A. 2008, in Society of Photo-Optical Instrumentation Engineers
  (SPIE) Conference Series, vol. 7020 of Society of Photo-Optical
  Instrumentation Engineers (SPIE) Conference Series (http://arxiv.org/abs/0806.4888).

%\bibitem[{{Wedemeyer} {et~al.}({2015a}){Wedemeyer}, {Bastian}, \& {SSALMON
%  group}}]{ssalmon_ssrv15}
%{Wedemeyer}, S., {Bastian}, T.~S., \& {SSALMON group}. {2015a}, submitted to Space Science Reviews
%%
%\bibitem[{{Wedemeyer} {et~al.}({2015b}){Wedemeyer}, {Bastian}, \& {SSALMON
%  group}}]{ssalmon_espm15}
%{Wedemeyer}, S., {Bastian}, T.~S., \& {SSALMON group}. {2015b}, submitted to Advances in Solar Physics
%%
%\end{thebibliography}
%

\end{document}